\documentclass[article,amssymb,amsmath]{revtex4}
   
\usepackage{pifont}
\usepackage[version=4]{mhchem}
\usepackage{array}
\usepackage{graphicx}
\usepackage{subfigure}
\usepackage{multirow}
\usepackage{footnote}
\usepackage{color}
\usepackage{chemmacros}
\usepackage{xcolor}
\usepackage{epstopdf}
\usepackage{natbib}
\usepackage{dcolumn}
\usepackage{bm}
\newcolumntype{L}[1]{>{\raggedright\let\newline\\\arraybackslash\hspace{0pt}}m{#1}}
\newcolumntype{C}[1]{>{\centering\let\newline\\\arraybackslash\hspace{0pt}}m{#1}}
\newcolumntype{R}[1]{>{\raggedleft\let\newline\\\arraybackslash\hspace{0pt}}m{#1}}
\newcommand{\xmark}{\ding{55}}%
\newcommand{\specialcell}[2][c]{%
  \begin{tabular}[#1]{@{}c@{}}#2\end{tabular}}

\def\be{\begin{equation}}
\def\ee{\end{equation}}
\def\bea{\begin{eqnarray}}
\def\eea{\end{eqnarray}}

\begin{document}

\title{Stochastic Lag Time in Nucleated Linear Self-Assembly}

\author{Nitin S. Tiwari}
\affiliation{Group Theory of Polymers and Soft Matter, Eindhoven University of Technology,
P.O. Box 513, 5600 MB Eindhoven, The Netherlands}
\author{Paul van der Schoot}
\affiliation{Group Theory of Polymers and Soft Matter, Eindhoven
University of Technology, P.O. Box 513, 5600 MB Eindhoven, The Netherlands}
\affiliation{Institute for Theoretical Physics, Utrecht University,
Leuvenlaan 4, 3584 CE Utrecht, The Netherlands}

\date{\today}

\begin{abstract}
Protein aggregation is of great importance in biology, e.g., in amyloid fibrillation. The aggregation processes that occur at the cellular scale must be highly stochastic in nature because of the statistical number fluctuations that arise on account of the small system size at the cellular scale. We study the nucleated reversible self-assembly of monomeric building blocks into polymer-like aggregates using the method of kinetic Monte Carlo. Kinetic Monte Carlo, being inherently stochastic, allows us to study the impact of fluctuations on the polymerisation reactions. One of the most important characteristic features in this kind of problem is the existence of a lag phase before self-assembly takes off, which is what we focus attention on. We study the associated lag time as a function of the system size and kinetic pathway. We find that the leading order stochastic contribution to the lag time before polymerisation commences is inversely proportional to the system volume for large-enough system size for all nine reaction pathways tested. Finite-size corrections to this do depend on the kinetic pathway.
\end{abstract}

\pacs{Valid PACS appear here}
\maketitle


\section{Introduction}
Protein aggregation is linked to neuro-degenerative afflictions such as Alzheimer's and Parkinson's disease. \cite{takalo} Another classic example of a linearly self-assembling system is the polymerization of actin fibrils and microtubules in living cells that help cells to control their mechanics. \cite{blanchoin} It is the importance of the aggregation of proteins in a wider biological context that has motivated many theoreticians and experimentalits to study equilibrium \cite{paul1, zlotnick} and kinetic aspects of it. \cite{cates_all, oosawa, cohen} Although decades of theoretical and experimental investigation has contributed to deeper understanding of how proteins self-assemble into polymer-like objects \textit{in vitro}, relatively little is known as to what happens \textit{in vivo}, that is, at the cellular scale. Almost all of the theoretical work is restricted to using deterministic rate equations to study the kinetics of linear self-assembly. Even experimentally, studying stochasticity in protein aggregation turns out to be challenging. \cite{ferron, tuomas_stochastic} This is because given the mesoscopic nature of the biology of the cell, the stochasticity is arguably mainly due to statistical number fluctuations of assemblies of a certain size. As cells are small in volume (pico litre to femto litre), the number of molecules aggregating is also relatively small, that is, on the scale of typical \textit{in vitro} experiments, causing fluctuations to play a significant role in the kinetics as well as in equilibrium. The latter expresses itself in large deviations from mean length distributions. \cite{wheeler}

The kinetics of self-assembly of protein monomers into polydisperse polymer-like objects, i.e., the time evolution of the polymer length distribution, can take place via a host of different molecular pathways. Examples include \textit{monomer addition and removal}, \textit{scission and recombination}, \textit{secondary nucleation} and so on, see Fig. 1. \cite{cates_review,cates_sr,cates_sr1,cates_eea} In principle, all these pathways are active during the time course of aggregation, but the predominance of one or more pathways may depend on the specific chemistry of a particular system. Due to the highly nonlinear nature of the dynamical equations that describe these pathways, it is extremely difficult to study the kinetics of even a single pathway analytically, let alone a combination of several of them. To study the \textit{stochastic} nature of self-assembly, one would be required to include the effects of thermal noise that might depend on a plethora of rate constants and thermodynamic parameters. \cite{cohen} Deducing the properties of the noise from microscopic rate equations, i.e., by making use of the fluctuation-dissipation theorem, is not trivial, again due to the complexity of the involvement of multiple molecular pathways. All of the above-mentioned challenges limits the applicability of analytical tools available to study the kinetics of protein aggregation. \cite{allen}

Given the challenges and limitations of the rate-equations approach, we study in this work the kinetics of nucleated linear self-assembly using the method of kinetic Monte Carlo, hereafter referred to as kMC. kMC is a computer simulation technique used to study Markov processes. \cite{gillespie} Protein aggregation can be cast in to the form of a Markov process, where a monomer becomes a dimer, a dimer becomes a trimer, and so on. This is exactly the same principle on which rate-equation approaches are based upon. The advantage of kMC is that it is inherently stochastic and hence we do not need to know the nature of thermal noise \textit{a priori}. In fact, the results obtained from this technique can be used to characterise this kind of noise and its dependence on the model parameters. Stochastic simulation techniques have been widely employed to study colloidal aggregation kinetics. \cite{thorn1,thorn2,odriozola1,odriozola2} Colloidal aggregation resembles protein self-assembly closely in terms of the governing rate equations. The characteristic feature of a lag phase present in nucleated self-assembly has also been found in computational studies of reversible colloidal aggregation. \cite{puertas}

In this paper we explore the stochastic nature of the kinetics of protein polymerization in small volumes, and study the effect of statistical number fluctuations. We study the aggregation kinetics for small volumes not for one pathway but for no fewer than nine of them discussed in more detail in Section II. The pathways we consider are \textit{end evaporation and addition} \cite{cates_eea}, \textit{scission and recombination} \cite{cates_sr}, \textit{secondary nucleation} \cite{cohen}, \textit{two-stage nucleation} \cite{tuomas_two_stage}, and combination thereof. For the \textit{scission and recombination} molecular pathway, the forward and backward rate constants in principle depend on the length of the polymer chains. Hence, to probe the impact of length-dependent rate constants, we use Hill's model. \cite{hill} Although Hill's model is valid only for long rigid rod like polymers, we neverthless use it to study if any kind of length dependence of the rate constants can alter the scaling of the lag time with system volume. To quantify the stochastic behavior of the polymerization kinetics, we focus specifically on the lag time associated with the time evolution of the polymerized mass fraction. We provide the simulation details in Section III. In Section IV, by doing an exhaustive study for various combinations of molecular pathway, we show that the lag time for the polymerized mass fraction is inversely proportional to the system size in the limit of large volumes albeit that there are corrections to that, that do depend on the specific assembly pathway. \cite{tuomas_stochastic, ferron, szabo, allen} Finally, in Section V, we conclude the article and discuss our main findings.

\section{Kinetic Pathways for Reversible Self-Assembly}

The molecular pathway determines the relevant time scales in reversible polymerization kinetics. \cite{cates_all, oosawa, cohen} Hence, if we are interested in the stochastic kinetics of the protein aggregation, it makes sense to study the system size dependence of a variety of molecular pathways. In this work we choose the most widely accepted and dominant kinetic pathways as far as the study of protein polymerization is concerned. \cite{cohen, tuomas_two_stage} We focus on (i) \textit{end evaporation and addition}, (ii) \textit{scission and recombination}, (iii) \textit{monomer-dependent secondary nucleation}, and (iv) \textit{two-stage nucleation}, and combinations of these pathways. In our description, all these reaction pathways have two things in common, which is that they all are nucleated with nucleus size $n_c$ and that they are \textit{reversible}. Primary nucleation is thought of as the coming together of a minimum of $n_c$ monomers to form the smallest stable polymer.

Of these pathways seemingly the most obvious to consider is that of \textit{end evaporation and addition}, given the linear structure of the aggregates of proteins. In it, growth or shrinkage of an aggregate is possible only by adding or removing a single monomer from either end. \cite{oosawa} Although \textit{end evaporation and addition} is the most plausible choice to describe the kinetics of the evolution of length distributions of aggregates, fragmentation and recombination of polymers must be important if the self-assembly is nucleated. For a nucleated self-assembling system, the \textit{end evaporation and addition} pathway requires each polymer formed to cross a nucleation barrier, which slows down the assembly process significantly. Instead, in \textit{scission and recombination} kinetics the polymers can bypass the nucleation barrier by breaking already existing filaments and creating new nucleation centers that then can grow. Hence, it is sensible to study scission-recombination in combination with end evaporation and addition, which is why this is one of the combined schemes that we investigate.

The forward and backward rate constants for all pathways could be length dependent, in particular for the scission-recombination pathway. Indeed, shorter polymers should recombine faster than longer ones if the reaction is diffusion limited, and longer ones plausibly have a higher probability of breaking within a certain amount of time than shorter ones. We do take into account the length dependence of the rate constants for the \textit{scission and recombination} pathway by invoking Hill's rate constants. \cite{hill} Hill derived the forward and backward rate constants in the limit of diffusion-limited aggregation that are valid only for long rigid rods. Hill's rate constants are not entirely consistent with the thermodynamics of nucleated self-assembly, and alternative ones have been derived in the context of the radical polymerization with more accurate length dependent rate constants these seem to suffer from the same problem. \cite{buback, nikitin} As our aim is to find out if any kind of length dependence of the rate constants alters the dominant scaling of the lag time as a function of system volume, we make an arbitrary choice of using Hill's model. We study both kinds of scission-recombination pathway, i.e., with and without length dependence of the rate constants.

The pathways of \textit{end evaporation and addition} and of \textit{scission and recombination} have proven to apply to the polymerisation of actin and tubulin filaments, but the time scales obtained from these pathways cannot explain other types of protein aggregation, such as that of sickle hemoglobin. \cite{eaton, ferron1, ferron2, ferron3} Sickle cell hemoglobin shows relatively rapid polymerization kinetics in comparison to predictions from \textit{end evaporation and addition} and \textit{scission and recombination} kinetics. To successfully explain the rapid polymerization observed in sickle cell hemoglobin, surface-catalysed or secondary nucleation has been included in several studies. \cite{ferron1, ferron2, ferron3} Secondary nucleation is also found to be relevant in the context of amyloid fibrillation. \cite{miranker} Recalling that primary nucleation concerns the conversion of a cluster of monomers into a stable polymer of shortest length, \cite{cremades, lee, pappu} in secondary nucleation a critical number $n_c$ of monomers may first form an unstable cluster that next transforms into a stable cluster, which in turn facilitates the elongation process.

Clearly, molecules may polymerize via a combination of molecular pathways. In this work we perform an exhaustive study of stochastic aggregation kinetics allowing for several combinations of pathway, and study how the resulting kinetics are affected by the system size. The pathways we study are listed in Table I. Mathematically, reversible polymerization can be represented by an infinite set of rate equations for all allowed chemical reactions. Indeed, a monomer reacts with a monomer to form a dimer, a dimer reacts with a monomer to form a trimer and so on. Let us first translate each pathway of Table I into a set of chemical reactions and write the corresponding rate equation for the polymers as
\begin{table}
\centering
\begin{tabular}{|C{2.8cm}|C{4.0cm}|L{7.5cm}|C{2.2cm}|}
 \hline
 Aggregation pathway & Reaction & Future state given the present state $(x,...,y_i,...)$  & Reaction rate \\ \hline
 primary nucleation & \begin{center} \ce{ $n_c$ $x$ <=>[k_n^+][k_n^-] $y_{n_c}$} \end{center} & \specialcell{Forward reaction: $(x-n_c,y_{n_c}+1,...,y_i,...)$ \\ Backward reaction: $(x+n_c,y_{n_c}-1,...,y_i,...)$} & \specialcell{$x^{n_c} k_n^+$ \\ $y_{n_c} k_n^-$} \\ \hline
 end evaporation and addition & \begin{center} \ce{$y_i$ + $x$ <=>[k_e^+][k_e^-] $y_{i+1}$} \end{center} & \specialcell{Forward reaction: $(x-1,...,y_i-1,y_{i+1}+1,...)$ \\ Backward reaction: $(x+1,...,y_{i}+1,y_{i+1}-1,...)$} & \specialcell{$2 x y_i k_e^+$ \\ $2 y_{i+1} k_e^-$} \\ \hline
 scission and recombination & \begin{center} \ce{$y_i$ + $y_j$ <=>[k_f^+][k_f^-] $y_{i+j}$} \end{center} & \specialcell{Forward reaction: $(...,y_i-1,...,y_j-1,..y_{i+1}+1)$ \\ Backward reaction: $(...,y_i+1,...,y_j+1,..y_{i+1}-1)$} & \specialcell{$y_i y_j k_f^+(i,j)$ \\ $y_{i+j} k_f^-(i,j)$} \\ \hline
 secondary nucleation & \specialcell{\ce{$i$ $y_i$ + $n_{\text{sec}}$ $x$ ->[k_{\text{sec}}] $y_{\text{sec}}$ + $i$ $y_i$}} & Forward reaction: $(x-n_{\text{sec}},...,y_{\text{sec}}+1...,y_i,...)$ & $i y_i x^{n_{\text{sec}}} k_{\text{sec}}$\\ \hline
 two-stage nucleation & \begin{center} \specialcell{\ce{$n_c$ $x$ <=>[k_{n}^+][k_{n}^-] $x_{n_c}$} \\ \ce{$x_{n_c}$ <=>[k_{c}^+][k_{c}^-] $y_{n_c}$}} \end{center}  & \specialcell{Forward reaction: $(x-n_c,x_{n_c}+1,...,y_i,...)$ \\ Backward reaction: $(x+n_c,x_{n_c}-1,...,y_i,...)$ \\ Forward reaction: $(x_{n_c}-1,y_{n_c}+1,...,y_i,...)$ \\ Backward reaction: $(x_{n_c}+1,y_{n_c}-1,...,y_i,...)$} & \specialcell{$x^{n_c} k_n^+$ \\ $x_{n_c} k_n^-$ \\ $x_{n_c} k_c^+$ \\ $y_{n_c} k_c^-$}\\ \hline
 \end{tabular}
 \caption{Possible molecular aggregation steps by which a polymer length distribution can change, considered in this work. Assuming the present state to be $(x,y_{n_c},...,y_i,...)$, where $x$ is the number of monomers and $y_i$ the number of polymers of length $n_c \le i \le \infty$, the states following the corresponding reactions are indicated. Notice that for secondary nucleation we do not have a backward reaction. This is because $y_{\text{sec}}$ is a polymer of size greater than the stable nucleus of size $n_c$, that then can disintegrate via monomer removal or scission. Also note that for two-stage nucleation we also have to track the evolution of unstable aggregate $x_{n_c}$.}
\end{table}
\bea
\frac{dy_i(t)}{dt} &=& k_n^+ x(t)^{n_c} \delta_{i,n_c} - k_n^- y_i(t) \delta_{i,n_c} \nonumber \\
&+& 2 k_e^+ x(t) y_{i-1}(t) - 2 k_e^+ x(t) y_{i}(t) + 2 k_e^- y_{i+1}(t) - 2 k_e^- y_i(t) \nonumber \\
&-& \sum_{j=n_c}^{i-n_c} k_f^-(i-j,j) y_i(t) + \sum_{j=n_c}^{\infty} k_f^-(i,j) y_{i+j}(t) + \sum_{k+l=i} k_f^+(k,l) y_{k}(t) y_{l}(t) - \sum_{j=n_c}^{\infty} k_f^+(i,j) y_i (t) y_j(t) \nonumber \\
&+& k_{\text{sec}} x(t)^{n_{\text{sec}}} \delta_{i,n_{\text{sec}}} \sum_{j=n_c}^{\infty} j y_j(t),
\eea
and that for the monomers as
\bea
\frac{dx(t)}{dt} &=& - \frac{d}{dt}\left( \sum_{i=n_c}^{\infty} i y_i \right),
\eea
which conserves the total amount of material in the solution. Here, $\delta_{i,n_c}$ is the Kronecker delta that obtains the value unity if $i=n_c$ and zero otherwise, and $x$, $y_i$ and $n_c$ denote the number of monomers, the number of polymers of degree of polymerization $i \ge n_c$ and the size of the critical nucleus, respectively, for a given system volume. The kinetic rate constants $k_n^+, k_n^-, k_e^+, k_e^-, k_f^+(i,j), k_f^-(i,j)$ and $k_{\text{sec}}$ are associated with the various molecular aggregation pathways listed in Table I. The rate constants associated with scission and recombination, i.e., $k_f^+(i,j)$ and $k_f^-(i,j)$ depend on  post-scission or pre-recombination polymer lengths, $i$ and $j$. In this work these rate constants are assumed to be length independent, in which case $k_f^+(i,j)=k_f^+$ and $k_f^-(i,j)=k_f^-$, except for a specific case of the Hill's length-dependent rate constants to be discussed below.

To use experimentally consistent units we assign molar units for the rate constants in Table I. However, as our simulation method is based on dealing with numbers of molecules rather than concentration, the reaction rate constants in our simulations are appropriately rescaled by the system volume to attain the dimensions of $s^{-1}$.

In Eq. (1), the first two terms describe \textit{primary nucleation}, while \textit{end evaporation and addition} contributes the next four terms, and \textit{scission and recombination} are described by the next four terms. The last term is a consequence of secondary nucleation. The factors of two in Eq. (1) accounts for the fact that each linear polymer has two ends. Eqs. (1) and (2)  describe the time evolution of the length distribution when the nucleation mechanism is straightforward primary nucleation. To describe the kinetics of aggregation pathways involving two-stage nucleation, we also have to consider the dynamics of unstable aggregates of size $n_c$. The resulting equations are in that case slightly different, where the number of unstable nuclei of $n_c$ monomers $x_{n_c}$ obeys
\bea
\frac{d x_{n_c}}{dt} &=& k_n^+ x(t)^{n_c} - k_n^- x_{n_c}(t) - k_c^+ x_{n_c}(t) + k_c^- y_{n_c}(t),
\eea
and the polymers follow
\bea
\frac{dy_i(t)}{dt} &=& k_c^+ x_{n_c}(t) \delta_{i,n_c} - k_c^- y_i(t) \delta_{i,n_c} \nonumber \\
&+& 2 k_e^+ x(t) y_{i-1}(t) - 2 k_e^+ x(t) y_{i}(t) + 2 k_e^- y_{i+1}(t) - 2 k_e^- y_i(t) \nonumber \\
&-& \sum_{j=n_c}^{i-n_c} k_f^-(i-j,j) y_i(t) + \sum_{j=n_c}^{\infty} k_f^-(i,j) y_{i+j}(t) + \sum_{j=n_c} k_f^+(i-j,j) y_{i-j}(t) y_{j}(t) - \sum_{j=n_c}^{\infty} k_f^+(i,j) y_i (t) y_j(t) \nonumber \\
&+& k_{\text{sec}} x(t)^{n_{\text{sec}}} \delta_{i,n_{\text{sec}}} \sum_{j=n_c}^{\infty} j y_j(t),
\eea
Mass conservation requires again that
\bea
\frac{dx(t)}{dt} &=& - \frac{d}{dt}\left( \sum_{i=n_c}^{\infty} i y_i \right) - n_c \frac{dx_{n_c}(t)}{dt}.
\eea
Here, $k_c^+$ and $k_c^-$ are forward and backward rates for unstable aggregation formation and its dissociation back to monomers, and all other rate constants have the same meaning as before.

As remarked above, the rate constants associated with the scission and recombination of polymers could depend on the length of the polymers. The length dependence of the scission and recombination rate constants $k_f^+$ and $k_f^-$ that we consider in this work were derived by Hill in the diffusion-limited aggregation regime. \cite{hill} They obey
\bea
k_f^-(i,j) &=& \frac{k_f^- (ij)^{n-1} (i \ln j + j \ln i)}{(i+j)^{n+1}},
\eea
for the backward rates, and
\bea
k_f^+(i,j) &=& \frac{k_f^+ (i \ln j + j \ln i)}{ij(i+j)},
\eea
for the forward rates, where $i \ge n_c$ and $j \ge n_c$ are the degrees of polymerization of the filaments, either recombining resulting into a filament of length $i+j$ or a filament of length $i+j$ fragmenting into two polymers, $n$ denotes the number of degrees of freedom of each polymer contributing to the diffusive transport of particles required to merge or separate two polymers. \cite{hill} These might include rotation, translation or even flexing of the polymer chains. Hong and Yong find in their work the value of $n \text{to apply} 1 \sim 3$ for several amyloid fiber systems. \cite{hong} 

For our purposes of qualitative study, we choose a value of $n=2$. The Hill's rate constant for scission is a bell-shaped curve peaked in the center, i.e., the polymer has largest probability of breaking around the middle, while the recombination rate decreases monotonically as a function of the size of the polymers engaged in recombining. It should be emphasized that Hill's model incorrectly predicts the length distribution for long times, and is anyway strictly applicable only for rigid rod-like polymer chains of $i,j \gg 1$. This then of course limits the applicability of Hill's model. We choose to ignore both caveats as we exclusively focus on the lag phase, i.e., the early time kinetics where the chains can already be very long sufficiently cooperative self-assembly.

The reaction rate equations Eqs. (1)-(5) completely characterize the time evolution of the length distribution for all aggregation pathways listed in Table I and for our choice of reaction rate constants. However, in this paper we study the self-assembly kinetics not by evaluating the rate equations, which are deterministic, but by means of the method of kinetic Monte Carlo applied to the reactions listed in Table I. This  simulation method, developed by Gillespie, has the advantage of being inherently stochsatic and has been applied extensively in the context of the kinetics of chemical reactions and of aggregation processes. \cite{mavelli, poschel} Fundamentally, the algorithm relies on two ingredients: (i) given the current micro-state of the system, choose the transition that takes the system to the next possible micro-state under the assumption of Markovian dynamics, and (ii) calculate the time for transition for the next micro-state, see also Appendix A. \cite{code_request}

To study the large number of combination of pathways we are interested in, we simply  switch on or off the desired pathways by making their rate constants non-zero or zero, respectively. We perform a consistency check by comparing the results of our stochastic simulations with predictions that we obtain using the deterministic Eqn. (1). We do this for one particular combination of three pathways, by obtaining closed form dynamical equations for the first two moments of the full distribution, i.e., the number of polymers and the polymerized monomeric mass. We focus on the combination of molecular pathways consisting of \textit{primary nucleation}, \textit{end evaporation and addition} and \textit{scission and recombination}, and presume length independent rate constants. The resulting moment equations we solve numerically and compare with our Monte Carlo simulations for large enough system size, where our simulations should produce deterministic predictions. As expected, our Monte Carlo simulations for $n_c=2$ are in quantitative agreement with the deterministic moment equations. This confirms the correct implementation of most of the individual reaction schemes. Details are presented in Appendix B.

Our stochastic simulations produce the time evolution of the full length distribution. Practically, it makes sense to focus on one or more moments of the length distribution, such as the polymerized mass, the polydispersity index and the number of ``living'' polymers. However, the latter two quantities are extremely difficult to probe experimentally, in particular as a function of time. Hence, in this work we exclusively focus on the first moment of the polymer length distribution excluding the inactive monomers. This is proportional to the polymerized mass fraction that is primarily probed in experiments and that is equal to the polymerized mass divided by the total monomer mass present in the solution.

In equilibrium the polymerized mass fraction that from now on we denote $f$, obeys $f=1-1/X$ if $X \ge 1$ and $f = 0$ if $X \le 1$, provided the polymerisation is sufficiently cooperative. Here, $X=C k_e^+/k_e^-$ is sometimes called a mass action variable, \cite{paul1} with $C$ the total monomer concentration and $k_e^+$ and $k_e^-$ the forward and backward rate constants for the \textit{end evaporation and addition} pathway. \cite{cates_eea} Notice that the rate constants for other pathways do not influence the equilibrium polymerized mass fraction. The reason is that nucleated reversible self-assembly can be seen as involving two components: (i) active polymers consisting of $n_c$ or more monomers and (ii) inactive monomers that can turn into active polymers. Upon incorporation of the \textit{scission and recombination} pathway the monomeric mass present in polymers remains unaffected. Indeed, for $i, j \ge n_c$ we cannot allow \textit{scission and recombination} kinetics to deplete or add to the monomer pool. Although the total number of monomers in the polymeric state is not influenced by this pathway, the polymers do tend to become shorter if $k_f^+/k_f^- \ge 1$, i.e, if the \textit{scission and recombination} pathway is switched on keeping all other reaction rates fixed. This implies that access to at least two moments of the full length distribution, such as the mean degree of polymerization and the polymerized mass fraction, is needed to ascertain the presence or absence of the \textit{scission and recombination} pathway in any experiment.
\begin{figure*}
\begin{center}
\includegraphics[width=6.5in]{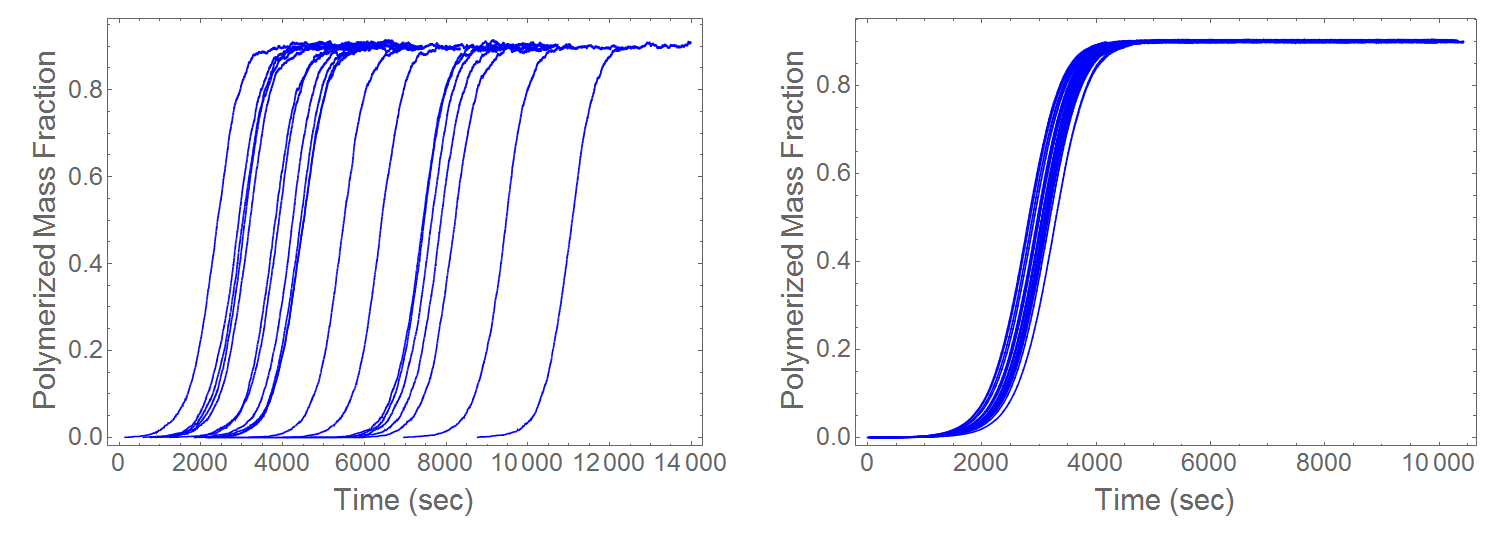}
\end{center}
\caption{Representative stochastic trajectories obtained from 20 different computer experiments performed for the set of parameters given in Table II, for system size (a) $V$=0.67 pL and (b) $V$=30 pL, with total monomer concentration of 10 $\mu M$ and a critical concentration of 1$\mu M$. Polymerization curves are shown for a combined molecular pathway with \textit{primary nucleation} (with nucleus size $n_c=2$), \textit{end evaporation and addition}, and \textit{scission and recombination}. The initial condition for the simulations is $y_i=0$ for all $n_c \le i \le  \infty$, i.e., only monomers are present at time $t=0$. The polymerization curves saturates at the polymerized mass fraction of 0.9 which is in agreement with the law of mass action.}
\end{figure*}

\section{System Size Dependence of The Lag Time}
In this section, we quantify the stochasticity of the polymerization kinetics by studying one key feature of linear self-assembly, known as the \textit{lag time} in the polymerized mass fraction. This is the most widely studied feature of nucleated self-assembly, experimentally and theoretically. \cite{tuomas_review} To quantify the lag phase for our system of nucleated self-assembly, we use the conventional definition in which we identify the time at which the growth rate is the largest, estimate the tangent at that point and finally take its time intercept as the lag time. \cite{hellstrand} In our study this is not a trivial affair because data obtained from our stochastic simulations are inherently noisy and, hence, straightforwardly \textit{calculating} derivatives is not feasible. To remedy this, we first obtain by means of a regression analysis a fit curve to our polymerization data points, using a generalized logistic function that has the following form, \cite{ref_file_exchange, curve_fitting}
\begin{table*}
 \begin{tabular}{|C{8.9cm}|C{1.5cm} C{0.85cm} C{0.85cm} C{0.85cm} C{0.85cm} C{0.85cm} C{0.85cm} C{0.85cm} C{0.85cm}|}
 \hline
 {\multirow{2}{*}{kinetic pathways}} & \multicolumn{9}{ c| }{reaction constants} \\ \cline{2-10}
   & $k_n^+$ & $k_n^-$ & $k_e^+$ & $k_e^-$ & $k_f^+$ & $k_f^-$ & $k_{\text{sec}}^+$ & $k_{c}^+$ & $k_{c}^-$ \\
   & $[M^{n_c-1}/s]$ & $[1/s]$ & $[M/s]$ & $[1/s]$ & $[M/s]$ & $[1/s]$ & $[M/s]$ & $[1/s]$ & $[1/s]$ \\ \hline \hline
  \specialcell{end evaporation and addition \\ ($n_c=1$)} & $10^{-6}$ & $10^{-1}$ & $10^3$ & $10^{-3}$ & \xmark & \xmark & \xmark & \xmark & \xmark \\ \hline
  \specialcell{end evaporation and addition \\ + scission and recombination ($n_c=1$)} & $10^{-6}$ & $10^{-1}$ & $10^3$ & $10^{-3}$ & $10^3$ & $10^{-3}$ & \xmark & \xmark & \xmark \\ \hline
  \specialcell{end evaporation and addition \\ + scission and recombination (Hill) ($n_c=1$)} & $10^{-6}$ & $10^{-1}$ & $10^3$ & $10^{-3}$ & $10^3$ & $10^{-3}$ & \xmark & \xmark & \xmark \\ \hline
  \specialcell{ evaporation and addition \\ ($n_c=2$)} & $10^{-2}$ & $10^{-3}$ & $10^3$ & $10^{-3}$ & \xmark & \xmark & \xmark & \xmark & \xmark \\ \hline
  \specialcell{end evaporation and addition \\ + scission and recombination ($n_c=2$)} & $10^{-2}$ & $10^{-3}$ & $10^3$ & $10^{-3}$ & $10^3$ & $10^{-3}$ & \xmark & \xmark & \xmark \\ \hline
  \specialcell{end evaporation and addition \\ + scission and recombination (Hill) ($n_c=2$)} & $10^{-2}$ & $10^{-3}$ & $10^3$ & $10^{-3}$ & $10^3$ & $10^{-3}$ & \xmark & \xmark & \xmark \\ \hline
  \specialcell{secondary nucleation \\ + end evaporation and addition ($n_c=2$)} & $10^{-2}$ & $10^{-3}$ & $10^3$ & $10^{-3}$ & \xmark & \xmark & $10^{-2}$ & \xmark & \xmark \\ \hline
  \specialcell{secondary nucleation + end evaporation and addition \\ + scission and recombination ($n_c=2$)} & $10^{-2}$ & $10^{-3}$ & $10^3$ & $10^{-3}$ & $10^3$ & $10^{-3}$ & $10^{-1}$ & \xmark & \xmark \\ \hline
  \specialcell{two stage nucleation + end evaporation and addition \\ + scission and recombination ($n_c=2$)} & $10^{-2}$ & $10^{-3}$ & $10^3$ & $10^{-3}$ & $10^3$ & $10^{-4}$ & \xmark & $10^{-3}$ & $10^{-2}$ \\ \hline
 \end{tabular}
 \caption{The kinetic rate constants for each molecular aggregation pathways and their combinations considered in this work. The parameters are defined in the main text. The molar (M) and seconds (s) are the units for concentration and time, respectively.}
\end{table*}
\bea
z(t)=\frac{z(\infty)}{(1+Q e^{-\alpha \nu t})^{\frac{1}{\nu}}},
\eea
where z denotes the polymerized mass fraction, t is time and $Q$, $\nu$ and $\alpha$ are fitting parameters.  We use the generalized logistic function instead of a simple logistic function to account for the asymmetry of polymerization kinetics before and after the inflection point. For every run, we construct a smooth function using our fitting procedure, calculate the maximum growth rate and determine the polymerized mass fraction at that point to calculate the lag time.

The lag time for small system size is not a deterministic function of the system parameters. Rather, it is a stochastic variable with a certain probability distribution. To obtain the distribution function of lag times, we repeat our computer experiment 500 times for the same set of parameters given in Table II. We study the system size dependence of the lag time distribution by performing our in-silico experiments for various volumes ranging from 0.3 pL to 30 pL (1 pL = $10^{-15} \mathrm{m}^3$), which is typical volume range for microfluidic experiments and living cells. \cite{tuomas_stochastic} The rate constants for our computer simulation are collected in Table II. We choose rate constants arbitrarily, but do aim to clearly see the effect of each pathway under consideration and at the same time perform simulations within reasonable time. The nucleation rate constants $k_n^+$ and $k_n^-$ are chosen in such a way that the nucleation constant, i.e, the ratio $k_n^+/k_n^-$, is small enough to give rise to a distinct lag phase. At the same time, the ratio $k_n^+/k_n^-$ should be large enough to have feasibly small mean polymer length so as to speed up the simulation. The concentration of protein monomers for all of our simulations is 10 $\mu M$, and the critical polymerization concentration for all of our simulation is $1\mu M$. This critical concentration can be inferred from the equilibrium thermodynamic theory of nucleated linear self-assembly with \textit{end evaporation and addition}. As discussed in the previous section, the additional pathway of \textit{scission and recombination} does not alter the equilibrium between monomers and polymers. This implies that the critical concentration is independent of the pathways considered.

We choose 10 $\mu M$ total monomeric concentration with a motive to stay deep into the polymerized regime. As is well known, nucleated reversible self-assembly is a true phase transition in the limit of infinitely large nucleation free energy barrier and hence occurs in the limit $k_n^+/k_n^- \rightarrow 0$. \cite{wheeler} Although in our computer experiments we have a finite (but large) nucleation barrier, this can give rise to critical fluctuations that have little to do with the effect of system size. Because our aim is to study system-size dependence and not any critical behavior, studying self-assembly dynamics deeply into polymerized regime avoids encountering fluctuations originating from the latter.

By fixing the concentration of monomers at 10 times the critical value and that way making certain that the stochasticity in our simulations finds its origin in the number fluctuations in each molecular species, we vary the number of molecules in the system by changing the system volume only. In Fig. 1 we show representative stochastic trajectories for the time evolution of the polymerized mass fraction for a small volume (0.3 pL) and a relatively large volume (30 pL). Notice that almost all of the stochasticity is confined to the lag time, i.e., the polymerization curves are shifted while preserving their shape, including the maximum growth rate (the inflection point). Fig 1. shows that if the lag time is shifted appropriately, all the polymerization curves for one set of parameters collapse on to a universal curve. This indicates that the only feature that is different for different runs is the lag phase. 

In Fig. 2 we show the distribution of lag times for different system sizes. Note that the lag time distribution for small volumes is exponential with a cutoff for times smaller than a value that does not seem to depend on system size, and tends slowly towards a normal distribution for larger system sizes. We find this for all molecular pathways tested. The system size in our simulations refers to volume at fixed monomer concentration, so to the number of monomers in the system. The gradual shift from piece-wise exponential to normal distribution with increasing system size (number of particles) is seen for all pathways tested and is in qualitative agreement with experiments on sickle cell hemoglobin by Ferron et al.. \cite{ferron} In their work, instead of changing volume to change the number of molecules in the system of observation they change concentration of monomers. Because the stochasticity in the self-assembly kinetics at the mesoscale arises as a consequence of statistical number fluctuations, changing volume or concentration while keeping the other constant produces similar results. We stresss again that from Fig. 2 we conclude that the exponential distribution is not a truly exponential one, because below a timescale $\tau_{\mathrm{min}}$ the probability of $\tau_{\mathrm{lag}} < \tau_{\mathrm{min}}$ rapidly tends to zero. We will discuss the detailed implication of this in the next section.

\begin{figure*}
\begin{center}
\includegraphics[width=6.5in]{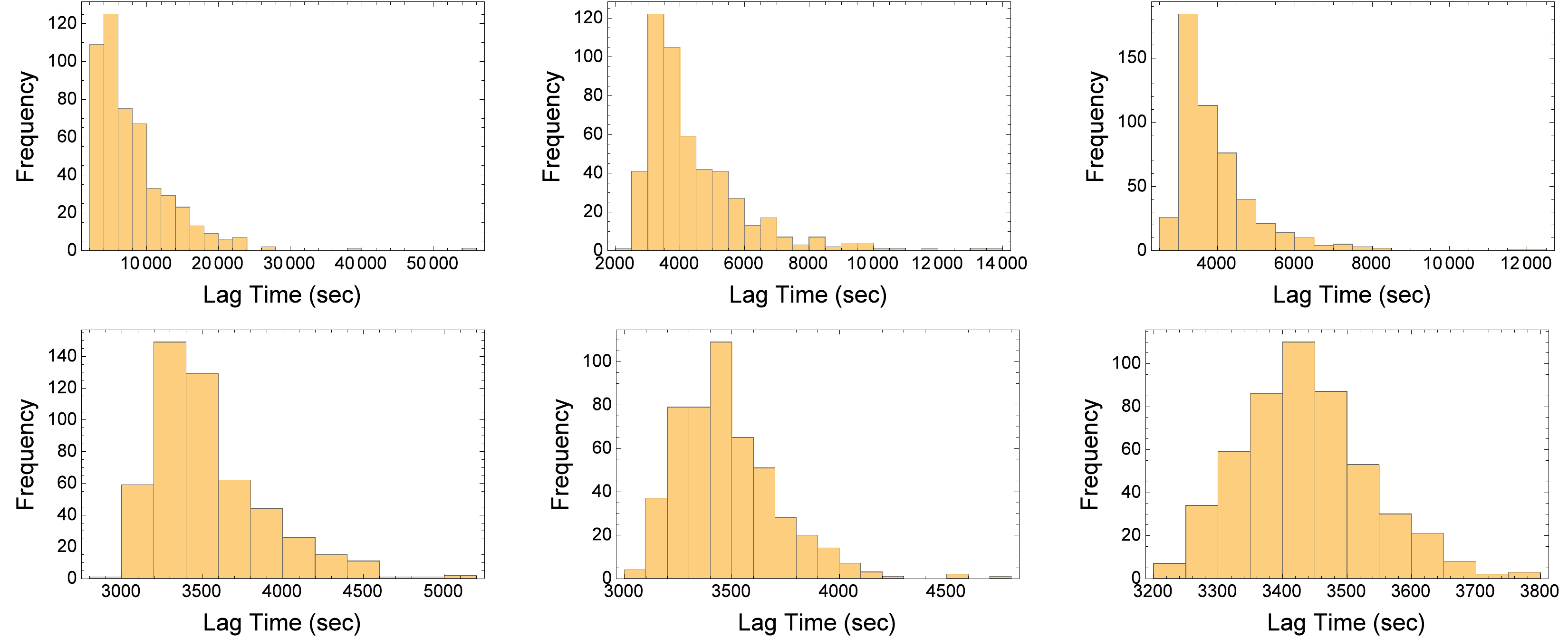}
\end{center}
\caption{The probability distributions of the lag time for systems of volume , $V=$ 0.30, 1.00, 1.67, 5.00, 8.33 and 30 pL, obtained by performing 500 runs for parameter values given in Table II. The data shown are for a combined molecular pathway with primary nucleation ($n_c=2$), end evaporation and addition, and scission and recombination. The simulations are performed under the total monomer concentration of 10 $\mu M$, where the critical concentration for polymerization is 1 $\mu M$. Although we only show the lag time distribution for one specific pathway, all the other combination of aggregation pathways listed in Table II also show similar qualitative gradual shift from exponential to gaussian with an increasing volume.}
\end{figure*}
From Fig. 2 we read off that both the mean lag time and its variance decreases with increasing system volume. We calculate the mean of the distribution of lag times and show in Fig. 3 the mean lag time as a function of inverse system volume $V^{-1}$. For generality we assume a power law dependence, i.e., $\tau_{\mathrm{lag}} = \tau_{\mathrm{lag}}^{\infty} + c/V^{\alpha}$, where $\tau_{\mathrm{lag}}^{\infty}$ is the infinite-volume, deterministic lag time, $c$ is a constant and $\alpha$ a power. After fitting with this power law, we obtain for the exponent $\alpha$ a value close to but not exactly equal to unity. The deviation of exponent $\alpha$ from unity is small and suggests a logarithmic correction from the universal law of $\tau_{\mathrm{lag}} \sim \tau_{\mathrm{lag}}^{\infty} + c V^{-1}$, i.e., $\tau_{\mathrm{lag}} \sim \tau_{\mathrm{lag}}^{\infty} + c V^{-1} (1- \delta \ln V)$, where $\delta$ is deviation of the exponent $\alpha$ from a value of unity. From Fig. 3, we conclude that to leading order the lag time is inversely proportional to the system volume. The power law of $\tau_{\mathrm{lag}} - \tau_{\mathrm{lag}}^{\infty} \propto 1/V$ is reached only for large volumes and is universal. The value of $\delta$ is not universal and depends on the reaction pathway. See Table III. Table III shows the exponent $\delta$ for all combinations of pathways listed in Table II. From the values listed in Table III, we confirm that the deviation $\delta$ is indeed pathway dependent. Hence, although the inverse system volume dependence of the lag time is a universal feature of all the studied combination of aggregation pathways, the deviation $\delta$ is non-universal. 

Another interesting cumulant of the lag time distribution is the variance of the lag time. It can be seen from error bars in Fig. 3 that the variance decreases with the system volume. Further analysis shows that the variance of the lag time distribution varies as $V^{-\beta}$, with $\beta$ strongly dependent on the pathway combination. For \textit{end evaporation and addition} the variance scales as $V^{-1/2}$, whereas addition the pathway of \textit{scission and recombination} changes this to $V^{-1}$ to the leading order albeit with an error as large as 50 percent. Indeed, the exponent of the variance of the lag time distribution is expected to be more erratic than the mean lag time, as higher moments are more sensitive to fluctuations. Given that this is the case, we cannot with absolute certainity comment on the scaling of variance with system volume.

Notice from Fig. 3 that different molecular aggregation indeed have different deterministic lag times $\tau_{\mathrm{lag}}^{\infty}$. To explain this, consider Fig. 2a for $n_c=1$, where among three sets of pathway shown, the \textit{end evaporation and addition} has the largest deterministic lag time. This is because, for every single polymer created, the system has to cross the nucleation barrier, thereby slowing down the assembly. If the \textit{scission and recombination} pathway with length independent rate constants is introduced, this allows the assembly to bypass the nucleation barrier by breaking already existing polymers and hence speed up the assembly process. This pathway has lowest deterministic lag time. By taking into account Hill's length dependent rate constants, we suppress the scission of smaller polymers as well as the recombination of longer ones. This, in turn, suppresses to some extent the energetically favourable creation of new polymers from already existing ones, and hence increases the assembly time. It is for this reason that the deterministic lag time for Hill's length dependent scission and recombination pathway falls between the \textit{end evaporation and addition} and the polymer length independent \textit{scission and recombination} pathways. The same reasoning holds true for bimolecular nucleation, so $n_c=2$. Fig. 2b and 2c shows that, not surprisingly, secondary nucleation also speeds up the assembly process. This is to be expected for secondary nucleation also allows the system to bypass the primary nucleation barrier by catalysing already formed polymers. The deterministic lag times for $n_c=1$ are different from those where $n_c=2$. This is because the lag time has a power law dependence on the initial polymerised mass fraction with the exponent being a function of the size of the critical nucleus $n_c$. \cite{cohen} The power law exponent also depends on the combination of pathway.

\begin{figure*}
\begin{center}
\includegraphics[width=6.5in]{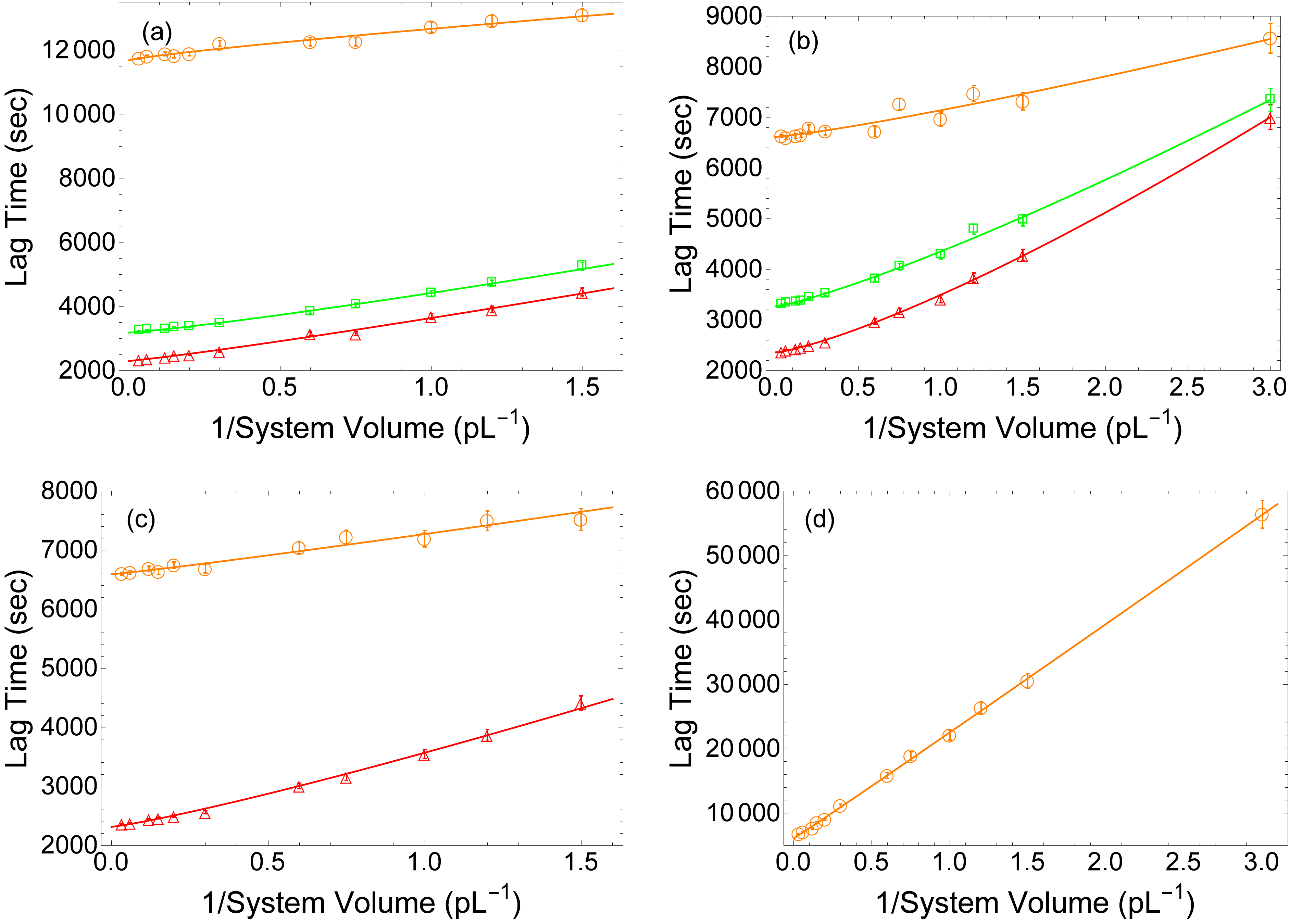}
\end{center}
\caption{Mean lag time as a function of the reciprocal system volume for various combination of pathways. (a) Primary nucleation (with nucleus size $n_c=1$) + end evaporation and addition (orange circle), primary nucleation ($n_c=1$) + end evaporation and addition + scission and recombination (red triangle) and primary nucleation ($n_c=1$) + end evaporation and addition + scission and recombination with Hill rate constants (green square). (b) Same as (a), except primary nucleation with $n_c=2$. (c) Primary nucleation ($n_c=2$) + secondary nucleation ($n_c=2$) + end evaporation and addition (orange circle) and  primary nucleation ($n_c=2$) + secondary nucleation ($n_c=2$) + end evaporation and addition + scission and recombination (red triangle). (d) Nucleation-conversion (two-stage nucleation) with ($n_c=2$) + end evaporation and addition + scission and recombination. All simulations are performed for total monomer concentration of 10 $\mu M$, where the critical polymerization concentration is 1 $\mu M$. Refer to Table II for the system parameters. The error bars indicate the variance of the lag time distribution.}
\end{figure*}
\newpage
\section{Discussion}
Naively, the universality of the leading order correction to the deterministic lag time originates from the requirement of a nucleation event. We infer this from Fig. 1, for the elongation phase following the lag phase is almost deterministic. To elucidate the actual cause of the existence of a lag time, we extract the first nucleus time, $\tau_{\mathrm{nuc}}$, from our computer simulations and show in Fig. 4 the probability distribution of this quantity for the combined molecular pathway with primary nucleation ($n_c=2$), \textit{end evaporation and addition}, and \textit{scission and recombination}. The other pathways paint qualitatively the same picture. From Fig. 4, we read off that unlike the lag time, the shape of the distribution function of the nucleation time is independent of the system size and is exponential without a small time cutoff. This is not surprising because the nucleation time is the time at which first nucleus is formed and, hence, can be seen as a first passage problem. \cite{chou} The first passage problem in our system is the transition of $n_c$ monomers into a stable nucleus. This nucleus is energetically unfavorable but, once a stable nucleus is formed, elongation can proceed.

Such first passage processes typically have a time scale associated with them that is exponentially distributed. \cite{vankampen} The independence of the shape of the nucleation time distribution on the system size hints at the circumstance that the lag time must be more than just a nucleation time. From the probability distribution from Fig. 4, we calculate the mean nucleation time, i.e., the average time to form first nucleus calculated from 500 runs, shown in Fig. 5. Following our analysis of the mean lag time, we assume the expectation value of the nucleation time $\tau_{\mathrm{nuc}}$ to depend on the system volume $V$ according to $\tau_{\mathrm{nuc}} = \tau_{\mathrm{nuc}}^{\infty} + c'/V^{\gamma}$ where $c'$ is a nucleation-mechanism dependent proportionality constant, $\tau_{\mathrm{nuc}}^{\infty}$ the nucleation time in the thermodynamic limit and $\gamma$ is the power law exponent for the stochastic nucleation time. Not surprisingly, Fig. 4 shows that the nucleation time remains unaltered for pathways affecting elongation mechanisms, i.e., \textit{end evaporation and addition} and \textit{scission and recombination}, and only depends on the primary nucleation constant $k_n^+/k_n^-$. Fig. 4b, 4c and 4d, have the same primary nucleation constant and same nucleus size of $n_c=2$, have similar volume dependence. Fig. 4a shows results for a nucleus size of $n_c=1$ and hence the rate of change of the nucleation time differs from the others. The nucleation time for $n_c=1$ is smaller than that for $n_c=2$, because of our choice of forward and backward rate constants.

The nucleation time turns out to be rather precisely linearly dependent on the system size, i.e., $\gamma = 1$ to within 1 to 8 percent. To explain this, we note that one particle has a first passage time, $\tau_p$, when the system crosses the nucleation barrier for the first time. For $N$ independent (uncorrelated) particles, the probability that one of them crosses the nucleation barrier will be $N$ times larger than the one particle case. Hence, the time scale of crossing the barrier will be inversely proportional to $N$, i.e., to the system volume. The same reasoning holds for the variance of the lag time distribution, because first passage processes generally result into exponentially distributed time scales. As is well known, the variance of the sum of $N$ exponentially distributed random variables with equal variance is simply the variance of one random variable divided by $N$. By the same token, $N$ particles crossing a nucleation barrier generate $N$ exponentially distributed first passage time scales. The probability distribution of the sum of these $N$ time scales is the so-called Gamma distribution, which in the limit $N \rightarrow \infty$ converges to Gaussian distribution, also see Fig. 2.

As remarked in previous section, we do observe such a linear dependence for the \textit{scission and recombination} pathway, but for the \textit{end evaporation and addition} there are large deviations from linearity. This is because the argument of $N$ independent particles crossing the nucleation barrier is not stricly applicable. The reason for this is that post-nucleation elongation of a polymer depletes the monomeric pool that, depending on the pathway, correlates the nucleation and elongation phases of the assembly. This results into a deviation from the linear dependence of the lag time on the system size, as observed. It implies that the nucleation time alone should scale as $N^{-1}$, which indeed can be seen from Fig. 5.

We notice from Figs. 3 and 5 that, for an infinitely large system, the deterministic nucleation time $\tau_{\mathrm{nuc}}^{\infty}$ is zero whilst the deterministic lag time $\tau_{\mathrm{lag}}^{\infty}$ is not zero. In fact, the distribution of nucleation times is exponential and that of the lag times is piece-wise, i.e., is essentially zero for small times. This observation strengthens our suggestion that the processes leading up to the existence of a lag time not only involves nucleation but also elongation. The part of the lag phase that involves elongation strongly depends on the elongation pathway considered, and hence does not have any of the universal features seen for the nucleation time. Different molecular pathways have different length distributions at the lag time and hence lack universality. In any event, the lag time defined the way described in Section I is an analytical way of quantifying a lag phase but it lacks any physical intuition. 

Indeed, the question what the length distribution is at the end of the lag phase cannot be generally answered. Hence, this may require us to redefine lag time and replace it by the elongation-pathway independent nucleation time and a pathway-dependent elongation time. How such a lag time is to be probed experimentally remain elusive though. The contribution of the elongation time to the lag time is caused by the circumstance the self-assembling system has to acquire a critical length distribution beyond which exponential growth occurs. The critical length distribution itself can be a stochastic variable with some distribution function. This adds further complexity to the problem of defining an elongation time. Although defining a sensible elongation time eludes us, we do emphasize that this is the most dominant time scale for the lag phase, at least in the thermodynamic limit. This can be seen once again from Figs. 3 and 5, where $\tau_{\mathrm{lag}}^{\infty} > 0$ whilst $\tau_{\mathrm{nuc}}^{\infty}=0$, hence confirming that for deterministic master equations, i.e., in the absence of any noise, the rate limiting step is elongation and not nucleation.
\begin{table}
 \begin{tabular}{|C{10cm}|C{2cm}|}
 \hline
  kinetic pathway & $\delta$ \\ \hline
  end evaporation and addition ($n_c=1$) & -0.17 \\ \hline
  \specialcell{end evaporation and addition \\ + scission and recombination ($n_c=1$)} & 0.12 \\ \hline
  \specialcell{end evaporation and addition \\ + scission and recombination (Hill) ($n_c=1$)} & 0.16 \\ \hline
  end evaporation and addition ($n_c=2$) & 0.19 \\ \hline
  \specialcell{end evaporation and addition \\ + scission and recombination ($n_c=2$)} & 0.28 \\ \hline
  \specialcell{end evaporation and addition \\ + scission and recombination (Hill) ($n_c=2$)} & 0.21 \\ \hline
  \specialcell{secondary nucleation \\ + end evaporation and addition ($n_c=2$)} & 0.09 \\ \hline
  \specialcell{secondary nucleation + end evaporation and addition \\ + scission and recombination ($n_c=2$)} & 0.16 \\ \hline
  \specialcell{two stage nucleation + end evaporation and addition \\ + scission and recombination ($n_c=2$)} & 0.02 \\ \hline
 \end{tabular}
 \caption{The aggregation pathways studied and their respective $\delta$, i.e., deviation from linearity of the lag time dependence on the system volume, $\tau_{\mathrm{lag}} - \tau_{\mathrm{lag}}^{\infty} \propto 1/V^{1+\delta}$.}
\end{table}
\begin{figure*}
\begin{center}
\includegraphics[width=6.5in]{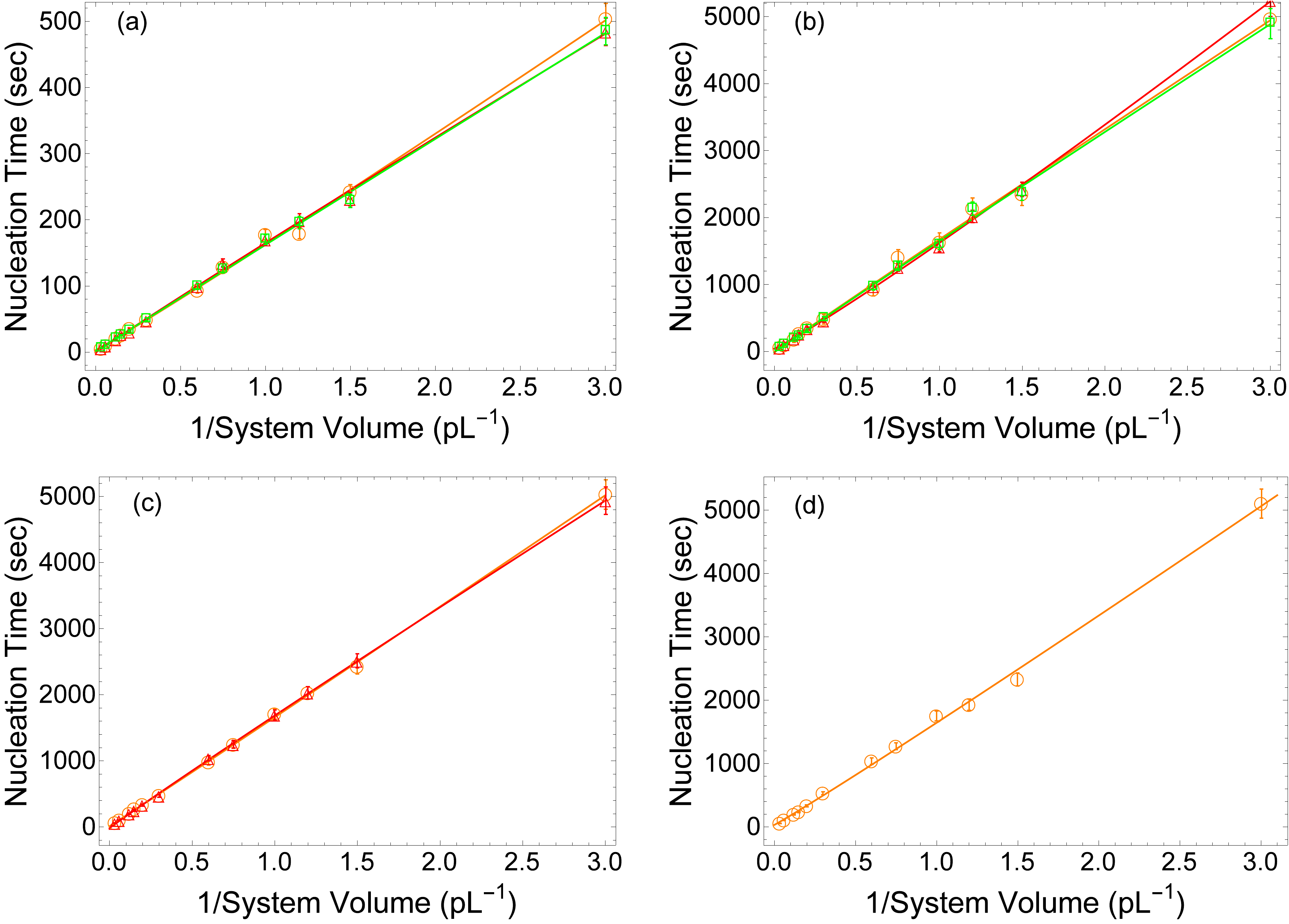}
\end{center}
\caption{The probability distribution of the nucleation time for system sizes of, $V=$ 0.30, 1.00, 1.67, 5.00, 8.33 and 30 pL, obtained by performing 500 computer experiments under the same set of parameters. The data shown for a combined molecular pathway with primary nucleation ($n_c=2$), end evaporation and addition, and scission and recombination. See also Table II for the values of the kinetic parameters. The total monomer concentration and the critical concentration is $10 \mu M$ and $1 \mu M$, respectively.}
\end{figure*}
\begin{figure*}
\begin{center}
\includegraphics[width=6.5in]{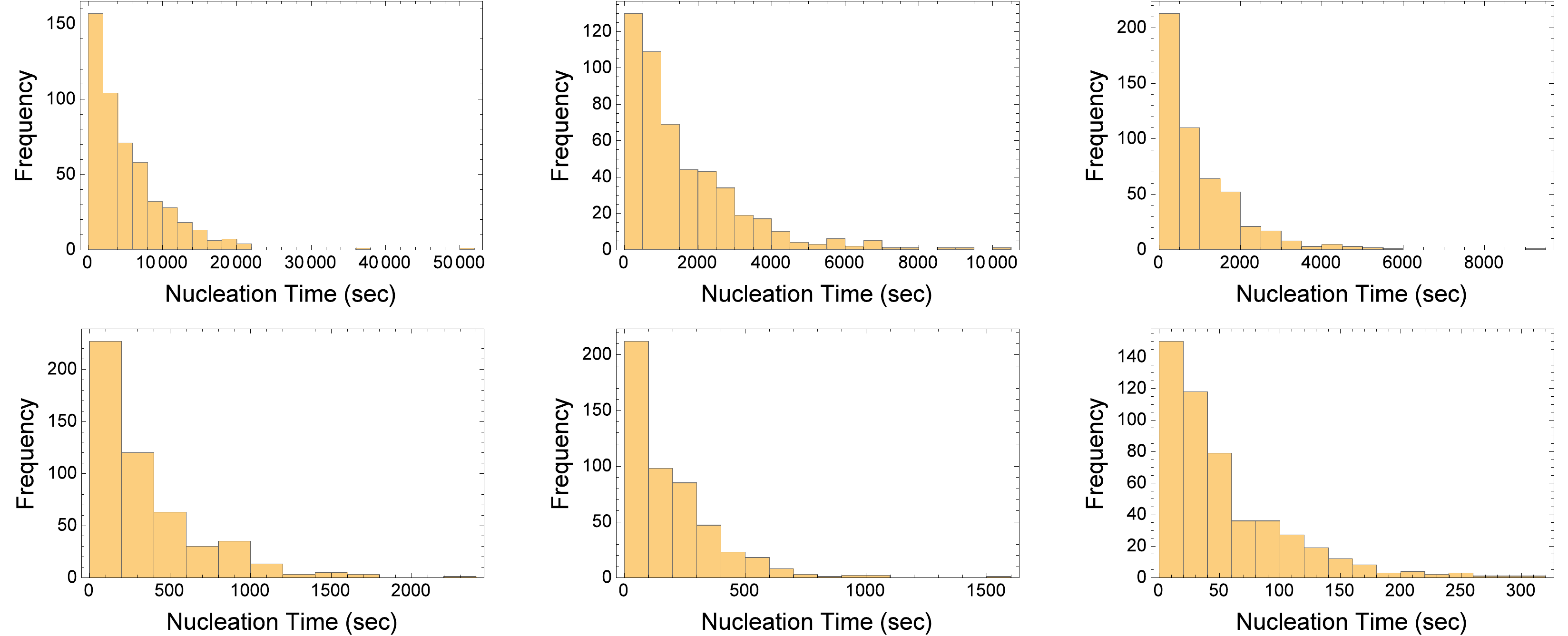}
\end{center}
\caption{Mean nucleation time as a function of the system volume for various combination of pathways. (a) Primary nucleation (with nucleus size $n_c=1$) + end evaporation and addition (orange circle), primary nucleation ($n_c=1$) + end evaporation and addition + scission and recombination (red triangle) and primary nucleation ($n_c=1$) + end evaporation and addition + scission and recombination with Hill rate constants (green square). (b) Same as (a), except primary nucleation with $n_c=2$. (c) Primary nucleation ($n_c=2$) + secondary nucleation ($n_c=2$) + end evaporation and addition (orange circle) and  primary nucleation ($n_c=2$) + secondary nucleation ($n_c=2$) + end evaporation and addition + scission and recombination (red triangle). (d) Nucleation-conversion (two-stage nucleation) with ($n_c=2$) + end evaporation and addition + scission and recombination. All simulations are performed for total monomer concentration of 10 $\mu M$, where the critical polymerization concentration is 1 $\mu M$. Refer to Table II for the system parameters.}
\end{figure*}
\section{Conclusions}
In this work we study by means of kinetic Monte Carlo simulation the stochastic nature of nucleated linearly self-assembling molecular building blocks in dilute solution. One of the models that we invoke is the thermodynamically consistent \textit{end evaporation and addition}, also known as the Oosawa model for self-assembly. \cite{oosawa} Another also includes \textit{scission and recombination}, with and without allowing for explicit length dependent rate constants. \cite{cates_review} We in addition allow for (i) monomolecular primary nucleation, (ii) bimolecular primary nucleation, (iii) secondary nucleation of monomers on already existing fibers and (iv) two-stage nucleation. \cite{tuomas_review} In combination, nine different sets of pathways are studied. We show that irrespective of the combination of pathways we study, to leading order the stochastic component of the lag time is inversely proportional to the system volume. This scaling remains unchanged even when Hill's length dependent rate constants, valid for rigid long polymer chains, are adapted in kinetic pathways. The first order correction that depends logarithmically on the volume turns out strongly pathway dependent. By comparing our lag time with the corresponding nucleation time to form the first nucleus, we show that for all tested pathways the stochastic component of the lag time must be a combination of a nucleation time and an elongation time. The nucleation time, unlike the lag time, is rather precisely inversely proportional to the system volume. We find it to be exponentially distributed for all system volumes, which is not the case for the lag time. The elongation time, on the other hand, strongly depends on whether the pathway involves only \textit{end evaporation and addition} or in addition \textit{scission and recombination} kinetics. This leads us to infer that a contribution from the elongation time is the cause of the non-universal correction to the leading order stochastic lag time. Finally, we find that in the thermodynamic limit the mean lag time is non-zero whilst the mean nucleation time seems to vanish. Consequently, for linearly self-assembling systems, the rate limiting step in the lag phase in that limit must be found in the elongation phase, not in the nucleation phase.

\section{Acknowledgment}
We thank Thomas Michaels (University of Cambridge) for stimulating discussions. This work was supported by the Nederlandse Organisatie voor Wetenschappelijk Onderzoek through Project No. 712.012.007.

\appendix

\section{Gillespie Algorithm for simulation of self-assembly}
We invoke the Gillespie algorithm to simulate our system of linearly self-assembling particles. \cite{gillespie} Let us consider a state of the system as $(x,y_2,y_3,...,y_i,...)$, where $x$ and $y_i$ are the numbers of monomers and of polymer of length $i$ respectively. The self-assembly as described by the molecular aggregation pathways shown in Table I, can be seen as a Markov process. In a Markov process the transition from the current state to the next state via one of the reactions shown in Table I depends on the current state of the system, and not on the states before that. As a first step towards implementing the Gillespie algorithm to study reaction kinetics is to list out all the possible reactions, given the current state of the system with their corresponding reaction rates. Given the current state of the system $(x,y_2,y_3,...,y_i,...)$, the Gillespie algorithm in essence requires two quantities: I) From all possible reactions listed in Table I, given the current state of the system, determine the next reaction that is going to take place in the time bracket from $t$ to $t+dt$. II) Calculate $dt$ the time that one of the reactions from Table I will happen for the first time.

To find next possible reaction, we first have to transform the reaction rates for each reaction of Table I into the corresponding probability. Let us define the rate to leave the present state, i.e., $R=\sum_{\alpha} R_{\alpha}$, where $R_{\alpha}$ is the reaction rate for each individual reaction $\alpha$. The probability of reaction $\alpha$ with reaction rate $R_{\alpha}$ is given by $P_{\alpha}=R_{\alpha}/R$. Next, we generate a uniformly distributed random number, $r_1$, in the interval $(0,1)$ and find the next possible reaction $\alpha$, such that, $\sum_{\beta=1}^{\alpha-1} P_{\beta} < r_1 < \sum_{\beta=1}^{\alpha} P_{\beta}$, where $\alpha$ is the reaction that is going to happen next. The second quantity, i.e., the time $dt$ for next reaction to happen turns out to be exponentially distributed, typical of first passage processes. \cite{vankampen} This then by simple transformation can be related to a uniform distribution. To calculate $dt$ we can then generate a uniformly distributed random number $r_2$, in the interval $(0,1)$ and using the relation $dt=\ln(1/r_2)/R$. This way, at every instance we know the next micro-state of the system and the time it will take for transition from the current to the next state of the system. This combination gives us the time evolution of the length distribution of the polymer self-assembly. By generating appropriate random numbers we also take into account the stochasticity coming from the mesoscopic number fluctuations. Hence, the Gillespie algorithm provides us with a tool to study stochastic kinetics arising from the reaction rate kinetics of self-assembling systems.

\section{Deterministic moment equations and comparison with Monte Carlo simulations} 
We obtain closed-form differential equations for the first two moments of a generalized pathway consisting of \textit{primary nucleation}, \textit{end evaporation and addition} and \textit{scission and recombination}. Similar equations have also been obtained previously by Michaels and Knowles. \cite{thomas_perturbative_techniques} Using Eq. (1) for the special case of length independent scission and recombination rate constants, we obtain for the polymers
\bea
\frac{dy_i(t)}{dt} &=& 2 k_e^+ x(t) y_{i-1}(t) - 2 k_e^+ x(t) y_{i}(t) + 2 k_e^- y_{i+1}(t) - 2 k_e^- y_i(t) - k_f^- (i- 2 n_c +1) y_i(t) \nonumber \\ 
 &+& 2 k_f^- \sum_{j=i+n_c}^{\infty} y_j(t) + k_f^+ \sum_{k+l=i} y_k(t) y_l(t) - 2 k_f^+ y_i (t) \sum_{j=n_c}^{\infty} y_j + k_n^+ x(t)^{n_c} \delta_{i,n_c},
\eea
and for the monomers
\bea
\frac{dx(t)}{dt} = -\frac{d}{dt}\left( \sum_{i=n_c}^{\infty} i y_i(t) \right).
\eea
Note the factor of $(i- 2 n_c +1)$ in the fifth term on the right hand side, accounts for the number of bonds allowed to break so that the fragmenting filaments are larger than $n_c$.
Next, we define first two moments of the full length distribution as
\bea
P(t)=\sum_{i=n_c}^{\infty} y_i(t) \quad \text{and} \quad M(t)=\sum_{i=n_c}^{\infty} i y_i(t),
\eea
where, $P(t)$ and $M(t)$ are the number of polymers and the polymerized monomeric mass respectively. Upon rearranging terms, we can write the time-evolution equation for $P(t)$ as
\bea
\frac{dP(t)}{dt} &=& 2 k_e^+ x(t) \sum_{i=n_c}^{\infty} \Big[ y_{i-1}(t)-y_i(t) \Big] + 2 k_e^- \sum_{i=n_c}^{\infty} \Big[ y_{i+1}(t)-y_i(t) \Big] - k_f^- \sum_{i=n_c}^{\infty} (i- 2 n_c +1) y_i (t) \\
& & + 2 k_f^- \sum_{i=n_c}^{\infty} \sum_{j=i+nc}^{\infty} y_j + k_f^+ \sum_{i=n_c}^{\infty} \sum_{k+l=i}^{\infty} y_k(t) y_l(t) - 2 k_f^+ \sum_{i=n_c}^{\infty} y_i (t) \sum_{j=n_c}^{\infty} y_j (t) + k_n x(t)^{n_c} \sum_{i=n_c}^{\infty} \delta_{i,n_c}. \nonumber
\eea
Note that the first and second term on the right hand side of Eq. (B4), coming from end evaporation and addition, cancel each other and hence do not contribute to the dynamics or equilibrium of first moment, $P$.

Next, we rewrite the third and the fourth term, coming from polymer scission, in terms of the theta function,
\bea
& & - \sum_{i=n_c}^{\infty} (i- 2 n_c +1) y_i(t) + 2 \sum_{i=n_c}^{\infty} \sum_{j=n_c}^{\infty} y_j(t) \Theta (i-j-n_c) \nonumber \\
&=& M(t)-(2 n_c -1) P(t).
\eea
The contribution from polymer recombination, represented by terms five and six, can be written as,
\bea
& &\sum_{i=n_c}^{\infty} \sum_{k+l=i}^{\infty} y_k(t) y_l(t) - 2 \sum_{i=n_c}^{\infty} y_i(t) \sum_{j=n_c}^{\infty} y_j(t) \nonumber \\
&=& \sum_{k=n_c}^{\infty} \sum_{l=n_c}^{\infty} y_k(t) y_l(t) - 2 \sum_{i=n_c}^{\infty} y_i(t) \sum_{j=n_c}^{\infty} y_j(t) \nonumber \\
&=& -P(t)^2.
\eea

The dynamical equation for total polymerized mass, $M(t)$, can be obtained by multiplying Eq. (B1) by $i$ and summing over it,
\bea
\frac{dM(t)}{dt} &=& 2 k_e^+ x(t) \sum_{i=n_c}^{\infty} i \Big[ y_{i-1}(t)-y_i(t) \Big] + 2 k_e^- \sum_{i=n_c}^{\infty} i \Big[ y_{i+1}(t)-y_i(t) \Big] - k_f^- \sum_{i=n_c}^{\infty} i (i- 2 n_c +1) y_i (t) \\
& & + 2 k_f^- \sum_{i=n_c}^{\infty} \sum_{j=i+nc}^{\infty} i y_j(t) + k_f^+ \sum_{i=n_c}^{\infty} \sum_{k+l=i}^{\infty} i y_k(t) y_l(t) - 2 k_f^+ \sum_{i=n_c}^{\infty} i y_i (t) \sum_{j=n_c}^{\infty} y_j (t) + k_n^+ x(t)^{n_c} \sum_{i=n_c}^{\infty} i \delta_{i,n_c}. \nonumber
\eea
Here, the first term, associated with elongation, can be simplified to
\bea
x(t) \sum_{i=n_c}^{\infty} i \Big[ y_{i-1}(t)-y_i(t) = \sum_{i=n_c-1}^{\infty} (i+1) y_i(t) - \sum_{i=n_c}^{\infty} i y_i(t) = P(t),
\eea
and the second (evaporation) term gives
\bea
\sum_{i=n_c}^{\infty} i \Big[ y_{i+1}(t)-y_i(t) \Big] = \sum_{i=n_c+1}^{\infty} (i-1) y_i (t)- \sum_{i=n_c}^{\infty} i y_i(t) = -P(t) - n_c y_{n_c}(t).
\eea
Here, we neglect the contribution, $n_c y_{n_c}(t)$, and obtain dynamical equation for second moment, $M(t)$. We can safely neglect the term $n_c y_{n_c}(t)$, because for sufficiently cooperative self-assembly, the number of nuclei are really small in comparison with the total number of polymers $P(t)$.

Using a similar algebraic manipulation as we did in Eqs. (B5) and (B6), we find that the terms originating from scission and recombination in Eq. (B7) vanish. This is to be expected, as we apply the scission and recombination pathway to the polymers only.  Again, $k_f^+$ and $k_f^-$ cannot influence the exchange of material between polymers and monomers. This leads us to the dynamical equations for the first two moments of the length distribution,
\bea
\frac{dP(t)}{dt} &=& - k_f^+ P(t)^2 + k_f^- \left( M(t)-(2 n_c -1) P(t) \right) + k_n x(t)^{n_c} \\
\frac{dM(t)}{dt} &=& 2 \big( x(t) k_e^+ P(t) - k_e^- P(t) \big) + n_c k_n^+ x(t)^{n_c}.
\eea
Analytical solution of the above equations has eluded us and hence we resort to  a numerical evaluation, the results of which we compare in Fig. (6) with those from our Monte Carlo simulations. For a comparison we took a system volume of $V=500$pL, which is very much larger than the volumes at which we see stochastic behavior. Stochastic behavior we find for our set of parameters to occur at volumes below approximately below $V=30$pL. See Fig. (1).
\begin{figure*}
\begin{center}
\includegraphics[width=6.5in]{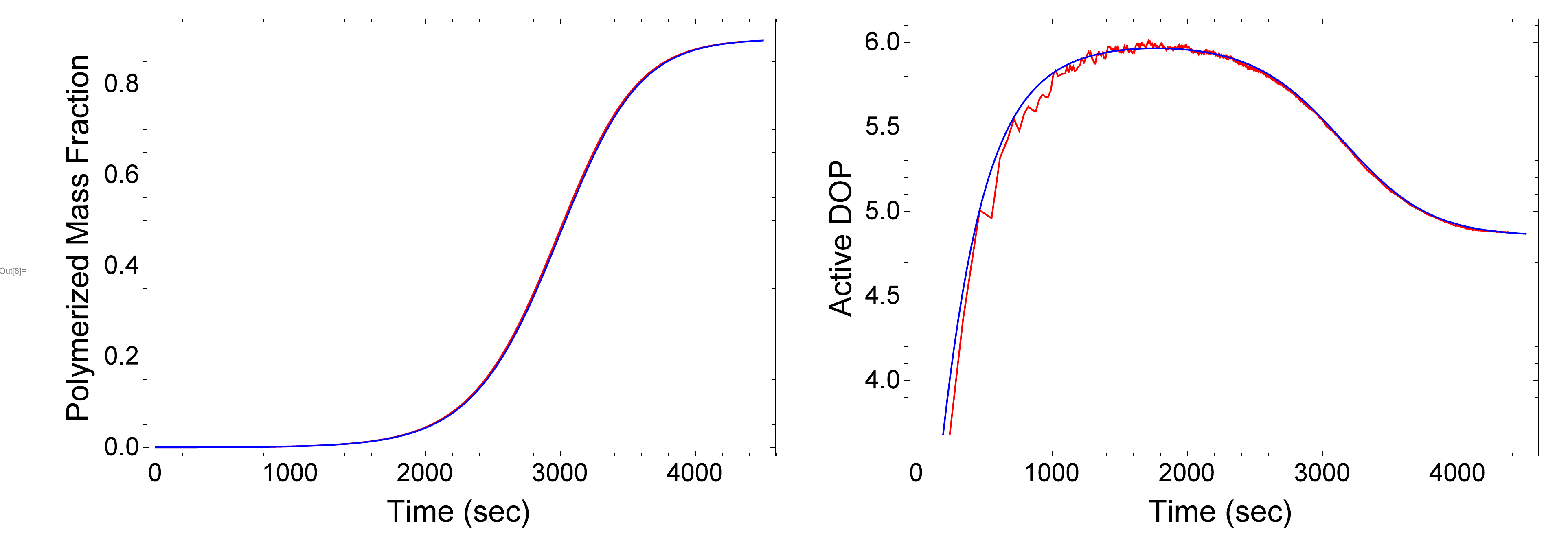}
\end{center}
\caption{Comparison of results obtained from our kinetic Monte Carlo simulations for a system volume of $V=500$pL, with a numerical solutions of the deterministic moment equations (B10) and (B11), valid in the infinite volume limit. Parameter settings as in Fig. 1. Left: the polymerized mass fraction, $M(t)/(M(t)+x(t))$, as a function of time. Right: The active degree of polymerization, $M(t)/P(t)$, as a function of time}
\end{figure*}


\begin{thebibliography}{}
\bibitem{code_request} The Python codes implementing the reaction schemes discussed in the main text, are available upon request.
\end{thebibliography}
\end{document}